# Selecting Sustainable Optimal Stock by Using Multi-Criteria Fuzzy Decision-Making Approaches Based on the Development of the Gordon Model: A case study of the Toronto Stock Exchange

Mohsen Mortazavi

1- Department of Computer Education and Instructional Technologies, Eastern Mediterranean University (EMU), Famagusta 99628, Cyprus.
2- Department of Information Technology, Islamic Azad University, Mahshahr Branch, Iran.
Email: mohsen.mortazavi.edu@gmail.com

**Abstract:**
Choosing the right stock portfolio with the highest efficiencies has always concerned accurate and legal investors. Investors have always been concerned about the accuracy and legitimacy of choosing the right stock portfolio with high efficiency. Therefore, this paper aims to determine the criteria for selecting an optimal stock portfolio with a high-efficiency ratio in the Toronto Stock Exchange using the integrated evaluation and decision-making trial laboratory (DEMATEL) model and Multi-Criteria Fuzzy decision-making approaches regarding the development of the Gordon model.

In the current study, results obtained using combined multi-criteria fuzzy decision-making approaches, the practical factors, the relative weight of dividends, discount rate, and dividend growth rate have been comprehensively illustrated using combined multi-criteria fuzzy decision-making approaches. A group of 10 experts with at least a ten-year of experience in the stock exchange field was formed to review the different and new aspects of the subject (portfolio selection) to decide the interaction between the group members and the exchange of attitudes and ideas regarding the criteria. The sequence of influence and effectiveness of the main criteria with DEMATEL has shown that the profitability criterion interacts most with other criteria. The criteria of managing methods and operations (MPO), market, risk, and growth criteria are ranked next in terms of interaction with other criteria.

This study concludes that regarding the model's appropriate and reliable validity in choosing the optimal stock portfolio, it is recommended that portfolio managers in companies, investment funds, and capital owners use the model to select stocks in the Toronto Stock Exchange optimally.

**Keywords:** Multi-Criteria Fuzzy, Gordon Model, Dividends, Discount Rate, Dividend Growth Rate, optimal stock portfolio.

# 1. Introduction

An important topic in financial literacy is choosing the optimal portfolio. Also, it aims to raise future returns and reduce investment risk. Identifying the factors in investor decision-making approach and their measurement and how they affect the selection and control of portfolios have been the two critical subjects for financial analysts. Generally, stock theories related to making portfolios can be broken into modern and postmodern groups. An important factor that can help investors optimize stocks is to pay heed to the criteria approved by financial experts. For investors, one of the critical issues is choosing stocks in the stock market [1], [2]. It is imperative to find ways to help investors choose stocks in these markets [3], [4]. One problem appears to choose the proper criterion for comparing different types of stocks [5]–[7]. Multi-criteria decision-making (MCDM) methods are suitable for studying various financial decision subjects.

Various factors affecting financial decisions (the appraisal of environment and goals), the intricacy of economic and business environments, and the subjective nature of plenty of financial decisions are the only characteristics of this multi-criteria decision modeling framework. They have proposed a survey of financial applications by implementing the Markov chains findings to proceed with an optimal portfolio. They considered a robust portfolio strategy and optimal portfolio formulas as a part of solving their problem, which was resolved by the dynamic programming principle (DPP) approach [8]. Stock selection and management are the two main areas of financial decision-making [9]–[11]. An essential factor that helps investors choose the right stock portfolio is to observe the criteria approved by experts and thinkers in finance. Investing in a portfolio is neither a linear nor a non-dimensional decision-making process [12]–[14]. A successful decision-maker examines any subject from different angles and uses several criteria jointly and simultaneously. Multidisciplinary decision-making approaches make it possible for investors to be acquainted with various practical aspects of market criteria. The investment process coherently requires a core analysis of the nature of decisions made in investment.

Here the activities relevant to the decision-making and the essential factors in the investment setting are observed and analyzed. Different factors and criteria are required to use multi-criteria methods. The multiple criteria are not the only major issue in the process, but the conflict between them and the allocation of weight to each indicator by each decision-maker is considered complicated dimensions [15]–[17]. According to what is stated above, it can be maintained that choosing the proper criteria for the evaluation of industries and types of stocks is to determine the weight and the relationship between them and rank the industry and stocks' main points in the process of making a decision [18]–[25].

Das et al. (2010) recommended the average return and its variance on the portfolio selection, which is generally presumed as the basis for choosing a portfolio investment by which securities' beliefs or predictions pursue similar probabilistic rules that random variables follow [26]–[29]. Many financial and economic theories are based on the concept by which rational people behave and regard all information in the decision-making process. Besides, researchers have found evidence that reflects irrational behavior and error repetition in human judgment & evaluation. Behavioral and psychological factors, such as mental accounting as risk factors, significantly affect decision-making [30].

A study by Chen et al. (2019) showed that people evaluate decisions independently and change the type of decision and the length of time and benefit [31]. Mental accounting explains that people mentally break their assets into distinct portfolios and want to hold each separately. In other words, exchanges are individually measured rather than evaluated repeatedly [26]. Decisions made by investors on the stock exchange to select stocks are based on factors such as earnings per share, liquidity, market trends (stock price trends), and the price-to-earnings ratio. The critical issue in the research is understanding the relative significance of each feature for investors. In other words, the main question in this research is raised as follows: " What are the practical factors in choosing stocks in the stock exchange market, and how do experts prioritize them?".

Shen et al. (2012) identified the criteria affecting share prices. According to a previous literature review, this study extracted the criteria affecting the three vital Gordon model sections [8]. These sections have been followed by the three main Gordon model criteria ( projected dividends, discount rate, and growth rate). It involves operating cash flow, industry outlook, earnings ratio, market beta, risk-free returns, and earnings growth rate [32].

Selecting a portfolio aims to evaluate a combination of securities from many available alternatives. Also, its objective is to expand investment returns for investors' output. Kazan and Uludağ (2014) presented that an investor must trade between the maximization of return and the minimization of risk. Thus, investors can expand the return for a risk level or concentrate on minimizing risk for a predetermined level of return. They calculated investment return and the value of securities' earnings. The risk is the value of variance [33]. The mean-variance (MV) model took variance in the returns and income from securities as its main inputs. Since then, several researchers have attempted to clarify the input data in the portfolio selection. Although a few methods, such as index models, have been successfully applied, plenty currently has limitations [33], [34]. The model is excessively considered fundamental since it neglects real-world issues related to investors, trading limitations, and portfolio size.

Kara et al. (2019) proposed a robust model optimizing the portfolio (where data was imported to the model from stocks markets) containing the "robust conditional value-at-risk under parallelepiped uncertainty" in order to find the robust optimal allocation of the portfolio numerically [35]. They compared their model with linear programming to see the risk-return analysis and price uncertainty [36], [37]. They have found that the proposed model can increase portfolio allocation stability and decrease portfolio risk [35].

Besides, all limitations in mathematics produce nonlinear mixed integer models that are very complex compared to basic ones. Though researchers have attempted to settle the issue through several approaches, such as cutting planes, interior point models, and decomposition, there is room for improvement [38]. Concerning risk and return, several studies have concentrated on MV models. MV models have been boosted to address real-world problems. However, some studies have ignored other significant issues in portfolio selection. There is, for instance, a controversy concerning the adequacy of regarding risk and return in selecting a portfolio; some additional criteria have been suggested by recent studies respecting additional criteria [39]. Therefore, the present study considers portfolio selection an MCDM problem.

*1-1 The criteria for Portfolio selection*

There are several criteria in the selection of a portfolio. These criteria appear to be different owing to the different concerns of managers, practitioners, researchers, and investors.

Although the portfolio criteria selection influences investors' final decisions, they were underexamined in the literature review. This is mainly because of the diversity and potential overlapping of criteria, making it difficult to recognize their differences.

Expected value (EV) is commonly applied to portfolio selection. Specifically, the methods of Tobin (1958), Markowitz (1952), and Sharpe (1963) are often used; however, these approaches faced numerous criticisms [40]–[42]. According to the research by Feldstein (1969) and Hakansson (1972), EV is applicable when the decision-makers expected utility is maximized, the utility function is quadratic, or the distribution probability of the return is expected [43]–[46].

Hurson and Ricci-Xella (2002) established an MCDM model with risk consideration applied to return rates, standard risk, and residual risk of portfolio selection.

Conflict criteria such as liquidity, risk, and return rates are often considered in a portfolio selection simultaneously [47]. For instance, Xidonas et al. (2009), and Ballestero et al. (2007), developed a multi-objective stochastic programming model with conflicting objective functions for selecting a portfolio. After filtering inefficiently, a decision table to consider multiple scenarios and choose portfolios has been provided using historical data [48], [49].

Liu et al. (2012) proposed an MCDM framework to choose typical stock portfolios, showing the suitability of MCDM approaches by applying transaction cost, return, skewness, and risk [50]. Meanwhile, potential criteria and sub-criteria were developed for selecting financial plans [51]. Table 1 summarizes portfolio selection criteria, showing sample references and applying factor and principal component analysis (PCA) to criteria development. Also, the table shows that whereas a few studies developed portfolio criteria selection based on literature reviews, a few have employed PCA or factor analysis to develop criteria. The study of the effectiveness of criteria is based on two principles. The first is the criteria studied based on previous research, and the second is the effective criteria based on financial ratios analysis. The following studies have been used to determine the four cluster criteria and their relationships. Having examined the components of the DuPont system, it can be shown that the company improves the shareholders' equity turnover by increasing total asset turnover (efficiency) and the leverage ratio (debt utilization). A company's financial risk is also dependent on its business risks. The investors can proceed or accept the higher financial risk if this parameter is negligible. According to Gordon's model, the dividend-to-earnings ratio affects the price-to-earnings ratio. Based on the affiliations of the CAPM model, the discount rate, one of the three main components of the Gordon model, is affected by market beta and risk-free interest rates. Also, the growth rate is a function of the dividend ratio that affects the price-to-income ratio. Indeed, different risks have profound impacts on market ratios.

The research conducted by Lee et al. (2009) stated that when a company predicts a steady cash flow for the future, thus, it pays more cash dividends. Thus, one of the parameters affecting this ratio is the growth of the company's revenues. Another parameter affecting dividend policy is financial leverage. So, immense financial leverage leads to more cash dividends being paid to shareholders. Companies with relatively high equity returns sell for several times their book value per share [52]. The identified criteria are classified into four main categories (in-network language in four clusters of profitability, market, growth, and risk clusters, as explained in more detail.

*1-2 Profitability cluster criteria:*
Return on Assets (ROA) is an asset utilization ratio that shows how effectively and efficiently the company uses its assets.

Return on Equity (ROE) is the net profit from ordinary shareholder rights units [53].

The net profit margin enables the company to generate income for each unit [54].

The operating profit margin enables the company to generate income for each unit.

Earnings per share (EPS) is the net income minus dividends of a preferred stock divided by the number of usual shares [34].

*1-3 Growth cluster criteria:*
Revenue Growth Rate is a firm's revenue change over a specified time.

The net profit growth rate (NPGR) is the degree of changes in a company's net profit over a specified time [55].

The earnings growth rate per share is defined as the degree of change in earnings per share of the company during a specified time [34].

A sustainable growth rate is a company's highest growth rate without increasing the leverage ratio and a capital increase.

*1-4 Risk cluster criteria:*
Business risk includes the company's revenue uncertainty since it has changed industry conditions. It is calculated by clustering the standard operating profit deviation to its average.

Financial risk is expressed as shareholders' uncertainty and return on equity risk. It corresponds to the company using other finance means with fixed liabilities (such as bonds) and its business risk.

Systematic risk (β) is the variability return on total securities. A direct relationship exists between general changes and market or general economic developments.

*1-5 Market cluster criteria:*

Market value to book value (MV / BV); indicates that this ratio explains the cross-sectional stock returns distribution [56].

Price-to-income ratio (P / E) shows that shareholders expect to recover the value of their current investment by the next few years (assuming conditions are maintained).

*1-6 Dividend ratio (DPS / EPS):*

This criterion shows the ratio of profit-to-profit per share. Shareholders usually seek to receive more cash dividends; on the other hand, the company's management is ready to distribute fewer cash dividends to earn more profits in the future by investing in accumulated dividends. A comprehensive study by Arkan (2016) used a series of financial ratios to predict the companies' financial strengths and compare these criteria's correlation with actual stock returns. The financial ratios are broken into profitability, operational efficiency, liquidity, leverage, company outlook, and growth criteria [57]. Is this study based on the explanations mentioned above and points concerning the environmental conditions of the Toronto Stock Exchange and the characteristics of the industries and companies listed in it? How can a mechanism be designed while extracting appropriate indicators and the preferences of investors in the analysis of industries and stocks to be utilized to select the use of stock portfolio with the highest returns? Thus, this research examines and prioritizes the factors influencing investors' decisions regarding the selection of stocks on the stock exchange. Here, the purpose of the decision-making approach is to choose the best and most logical option from all possible alternatives. The most appropriate decision is based on the relevant criteria.

*1-7 Research Objectives*

The aim is to determine the criteria for selecting shares in the Toronto Stock Exchange. The criteria were selected based on the research literature, and in order to review the criteria extracted from the review of literature, a group of 10 experts who had at least ten years of experience in the stock exchange field was formed to review the different and new aspects of the subject (portfolio selection) is to decide the interaction between the group members and the exchange of attitudes and ideas regarding the criteria. Therefore, the criteria extracted from the literature review were provided to the experts as a checklist to discuss each criterion collectively. After forming an expert panel, the experts' opinions were compiled, and their desired amendments were made.

At this stage, the criteria from the literature review were compiled and modified with the viewpoints of experts, and the experts' criteria, which did not exist in the literature review, were added to those criteria. Finally, 15 criteria were approved in 5 main criteria and used to select the portfolio. Specific objectives are as follows:

Determining the impact and effectiveness of each adequate criterion on the stock selection in the Toronto Stock Exchange.

Determining the priority of influential factors in selecting stocks in the Toronto Stock Exchange.

Therefore, the Toronto stock exchange is used in the current study.

## 2. Materials and Methods
### 2-1 Fuzzy DEMATEL

DEMATEL is a cluster-based MCDM method displaying the relationship between the causes and effects among criteria. It is categorized as the following steps;

**Step 1:** A questionnaire was drawn based on the two square matrices. It has a pairwise criteria comparison.

**Step 2:** To systematically analyze interrelations among the criteria, 21 experts were invited by the tools of pairwise comparisons.

**Step 3:** In this stage, experts administered ten linguistic variables to reflect the degree of causality among proposed degrees of impediments. To define the degree of impediment impact, the linguistic variables and corresponding triangular fuzzy numbers are shown in Table 1.

Table 1. Linguistic variables for the degree of influence of the impediment.

| Linguistic | Crisp | Fuzzy |
|---|---|---|
| **No influence (NI)** | 0 | (0,0,0) |
| **Extremely low influence (ELI)** | 1 | (0,0,0.1) |
| **Very low influence (VLI)** | 2 | (0,0.1,0.2) |
| **Moderately low influence (MLI)** | 3 | (0.1,0.2,0.3) |
| **Low influence (LI)** | 4 | (0.2,0.3,0.4) |
| **Medium influence (MI)** | 5 | (0.3,0.4,0.5) |
| **High influence (HI)** | 6 | (0.4,0.5,0.6) |
| **Moderately high influence (MHI)** | 7 | (0.5,0.6,0.7) |
| **Very high influence (VLI)** | 8 | (0.6,0.7,0.8) |
| **Extremely high influence (ELI)** | 9 | (0.7,0.8,0.9) |
| **Very Extremely high influence (VELI)** | 10 | (0.8,0.9,1) |

Then, the linguistic variables are converted into the triangular fuzzy number. The initial direct-relation matrix for the k$_{th}$ expert was made as $(\tilde{X}^k = [\tilde{X}^k_{ij}]_{n \times n} \ k = 1, 2, \cdots, m; m = 30)$. n is defined as the number of criteria where $i, j = 1, 2, \cdots, n$.

Each element in the matrix $\tilde{X}^k$ i.e. $\tilde{X}^k_{ij} = (l^k_{ij}, m^k_{ij}, u^k_{ij})$ is a triangular fuzzy number that depicts the i$^{th}$ criteria degree, which impacts the j$^{th}$ criteria when $i \neq j$. It is equal to (0,0,0) $when\ i = j$.

**Step 4:** To achieve fuzzy triangular numbers, Eq. (1) aggregates the direct-relation matrix
$$\tilde{A} = [\tilde{a}_{ij}]_{n \times n} \ (i, j = 1, 2, \cdots, n). \ \tilde{a}_{ij} = \frac{1}{m}[x^1_{ij} + x^2_{ij} + \cdots + x^m_{ij}] \tag{1}$$
Where (+) denotes Chen's triangular fuzzy numbers in addition to the operator.

**Step 5:** $\tilde{G} = [\tilde{g}_{ij}]_{n \times n} \ (i, j = 1, 2, \cdots, n)$ would be normalized by the direct-relation matrix. It was applied to Eqs. (2)-(4).

By assuming the element aggregation to the direct-relation matrix, $\tilde{A}$ is $\tilde{a}_{ij} = (l'_{ij}, m'_{ij}, u'_{ij}) \ (i, j = 1, 2, \cdots, n)$.

$$[\tilde{c}_i]_{n \times 1} = \left(\sum_{j=1}^n l'_{ij}, \sum_{j=1}^n m'_{ij}, \sum_{j=1}^n u'_{ij}\right) \quad i = 1, 2, \cdots, n \tag{2}$$

$$c = \max_{1 \leq i \leq n} \left(\sum_{j=1}^n u'_{ij}\right) \tag{3}$$

$$\tilde{G} = [\tilde{g}_{ij}]_{n \times n} = \left(\frac{\sum_{j=1}^n l'_{ij}}{c}, \frac{\sum_{j=1}^n m'_{ij}}{c}, \frac{\sum_{j=1}^n u'_{ij}}{c}\right) \quad i, j = 1, 2, \cdots, n \tag{4}$$

**Step 6:** The total direct-relation matrix $\tilde{S} = [\tilde{s}_{ij}]_{n \times n} \ (i, j = 1, 2, \cdots, n)$ where $\tilde{s}_{ij} = (l''_{ij}, m''_{ij}, u''_{ij})$ can be calculated by clustering $\tilde{G}$ and $\tilde{S}$ to three lower, middle, and upper elements of fuzzy triangular numbers in crisp matrices. Then, $\tilde{S}$ can be obtained by the utilization of Eqs. (5)-(7).

$$[l''_{ij}]_{n \times n} = [l'_{ij}]_{n \times n} \times \left([I]_{n \times n} - [l'_{ij}]_{n \times n}\right)^{-1} \tag{5}$$

$$[m''_{ij}]_{n \times n} = [m'_{ij}]_{n \times n} \times \left([I]_{n \times n} - [m'_{ij}]_{n \times n}\right)^{-1} \tag{6}$$

$$[u''_{ij}]_{n \times n} = [u'_{ij}]_{n \times n} \times \left([I]_{n \times n} - [u'_{ij}]_{n \times n}\right)^{-1} \quad (7)$$

Where I, is the identity matrix of order n.

**Step 7:** $\tilde{D} = [\tilde{d}_i]_{n \times 1}$, $(i = 1, 2, \cdots, n)$ and $\tilde{R} = [\tilde{r}_j]_{n \times 1}$ $(j = 1, 2, \cdots, n)$ The sum of the row and column is stamped as two vectors, respectively.

The addition of $\tilde{D}$ to $\tilde{R}$ $(\tilde{D} + \tilde{R})$ "Prominence" indicates the importance of each criterion.

$(\tilde{D} - \tilde{R})$ "relation" was obtained by subtracting $\tilde{D}$ from $\tilde{R}$.

Then, $D = [d_i]_{n \times 1}$, $R = [r_j]_{1 \times n}$, $\vec{D} + \vec{R}$ and, $\vec{D} - \vec{R}$ Vectors were defuzzified by Best Non-Performance (BNP) method.

When $i = j$, if $d_i > r_j \rightarrow d_i - r_j > 0$, then the criterion is a net cause; When $i = j$, if $d_i < r_j \rightarrow d_i - r_j < 0$, then the criterion is a net effect. $d_i$ Shows the sum of direct and indirect effects of criterion i on other criteria. $r_j$ indicates the sum of direct and indirect effects on criterion j.

## 3. Results

After extracting the primary definition and sub-criteria in the following, the interrelationship and their intensity of efficiency and effectiveness have also been measured. Then, all linguistic variables were changed to fuzzy triangular numbers, and using Eq. (1), the expert's opinions were combined (see Fig, 1-4). Finally, a crisp and fuzzy total direct-relation matrix and the normalized direct-relation matrix have also been calculated.

They are depicted in Tables 2 and 7, namely.

**Table 2.** The fuzzy total direct-relation matrix ($\tilde{S}$).

| S | | Profitability | Growth | Risk | Market | Macro |
|---|---|---|---|---|---|---|
| Main impediments | Abb | M01 | M02 | M03 | M04 | M05 |
| Profitability | M01 | (0.08,0.163,0.327) | (0.255,0.363,0.555) | (0.236,0.346,0.541) | (0.19,0.297,0.491) | (0,0.067,0.185) |
| Growth | M02 | (0.18,0.282,0.46) | (0.083,0.167,0.332) | (0.193,0.3,0.488) | (0.225,0.327,0.511) | (0,0.064,0.177) |
| Risk | M03 | (0.139,0.229,0.391) | (0.122,0.218,0.389) | (0.066,0.137,0.284) | (0.194,0.271,0.438) | (0,0.057,0.159) |
| Market | M04 | (0.212,0.317,0.502) | (0.234,0.344,0.538) | (0.226,0.337,0.532) | (0.094,0.178,0.35) | (0,0.067,0.186) |
| Macro | M05 | (0.286, 5,0.675) | (0.302,0.447,0.708) | (0.292,0.439,0.702) | (0.302,0.443,0.702) | (0,0.085,0.24) |

**Table 3.** The crisp total direct-relation matrix (s).

| S | Defuzzied | Profitability | Growth | Risk | Market | Marco |
|---|---|---|---|---|---|---|
| | | M01 | M02 | M03 | M04 | M05 |
| Profitability | M01 | 0.183 | 0.384 | 0.367 | 0.319 | 0.080 |
| Growth | M02 | 0.301 | 0.187 | 0.320 | 0.347 | 0.076 |
| Risk | M03 | 0.247 | 0.237 | 0.156 | 0.294 | 0.068 |
| Market | M04 | 0.337 | 0.365 | 0.358 | 0.200 | 0.080 |
| Macro | M05 | 0.453 | 0.476 | 0.468 | 0.472 | 0.102 |

**Table 4.** The sum of rows ($\tilde{D}$) and the sum of columns ($\tilde{R}$) for the fuzzy total-relation matrix and their corresponding crisp values.

| Abb. | Main Impediments | $\tilde{D}$ in triangular fuzzy form | $\tilde{R}$ in triangular fuzzy form | D in crisp form | R in crisp form |
|---|---|---|---|---|---|
| M01 | Profitability | (0.761,1.236,2.1) | (0.897,1.415,2.355) | 1.333 | 1.520 |
| M02 | Growth | (0.682,1.139,1.968) | (0.997,1.538,2.522) | 1.232 | 1.649 |
| M03 | Risk | (0.522,0.911,1.661) | (1.013,1.558,2.548) | 1.001 | 1.669 |
| M04 | Market | (0.766,1.242,2.109) | (1.006,1.517,2.493) | 1.340 | 1.633 |
| M05 | Macro | (1.183,1.838,3.027) | (0,0.339,0.947) | 1.972 | 0.406 |

**Table 5.** The "Relation" and "Prominence" values in the form of triangular crisp and fuzzy numbers

| Abb. | Main | $\tilde{D} + \tilde{R}$ | $\tilde{D} - \tilde{R}$ | $(\vec{D} + \vec{R})$ | $(\vec{D} - \vec{R})$ | Category |
|---|---|---|---|---|---|---|
| M01 | Profitability | (1.658,2.65,4.455) | (-0.137,-0.179,-0.255) | 2.853 | -0.750 | net effect |
| M02 | Growth | (1.679,2.677,4.49) | (-0.315,-0.399,-0.554) | 2.880 | -1.667 | net effect |
| M03 | Risk | (1.535,2.469,4.208) | (-0.492,-0.647,-0.887) | 2.670 | -2.672 | net effect |
| M04 | Market | (1.772,2.759,4.601) | (-0.24,-0.274,-0.384) | 2.973 | -1.172 | net effect |
| M05 | Marco | (1.183,2.178,3.975) | (1.183,1.499,2.08) | 2.378 | 6.261 | net cause |

**Table 6.** The sum of ($\tilde{D}$) and ($\tilde{R}$) for fuzzy total-relation matrix and their corresponding crisp values

| Abb. | Sub-impediments | $\tilde{D}$ in triangular fuzzy number form | $\tilde{R}$ in triangular fuzzy number form | D in crisp form | R in crisp form |
|---|---|---|---|---|---|
| **M011** | (ROA) return on assets | (0.557,1.099,2.224) | (0.899,1.551,2.91) | 1.245 | 1.728 |
| **M012** | (ROE) return on equity | (0.686,1.271,2.485) | (1.021,1.713,3.155) | 1.428 | 1.900 |
| **M013** | (EPS) earning per share | (0.44,0.944,1.988) | (0.618,1.178,2.344) | 1.079 | 1.330 |
| **M021** | revenue growth rate | (0.497,1.019,2.102) | (0.693,1.279,2.495) | 1.159 | 1.436 |
| **M022** | net profit growth rate | (0.46,0.969,2.026) | (0.773,1.383,2.656) | 1.106 | 1.549 |
| **M023** | sustainable growth rate | (0.445,0.95,1.997) | (0.586,1.135,2.279) | 1.086 | 1.284 |
| **M031** | economical risk | (0.509,1.035,2.126) | (0.75,1.353,2.61) | 1.176 | 1.516 |
| **M032** | financial risk | (0.466,0.977,2.037) | (0.804,1.425,2.718) | 1.114 | 1.593 |
| **M033** | systematic risk(beta) | (0.543,1.08,2.195) | (0.66,1.234,2.428) | 1.225 | 1.389 |
| **M041** | market value to price (MV/BV) value | (0.509,1.034,2.124) | (0.664,1.238,2.435) | 1.175 | 1.394 |
| **M042** | (P/E) price to income ratio | (0.498,1.021,2.104) | (0.639,1.207,2.387) | 1.161 | 1.360 |
| **M043** | (DPS/EPS) divided ratio | (0.465,0.946,1.991) | (0.658,1.231,2.424) | 1.087 | 1.386 |
| **M051** | exchange rate | (1.041,1.712,3.156) | (0.187,0.606,1.474) | 1.905 | 0.718 |
| **M052** | bank interest rate | (1.041,1.713,3.156) | (0.109,0.502,1.316) | 1.905 | 0.607 |
| **M053** | inflation rate | (1.003,1.692,3.124) | (0.098,0.427,1.203) | 1.878 | 0.539 |

Table 7. The "Relation" and "Prominence" values in the form of triangular fuzzy and crisp numbers

| Abb. | | $\widetilde{D} + \widetilde{R}$ | $\widetilde{D} - \widetilde{R}$ | $(\vec{D} + \vec{R})$ | $(\vec{D} - \vec{R})$ | status |
|---|---|---|---|---|---|---|
| M011 | (ROA)Return on assets | (1.456,2.65,5.134) | (-0.342,-0.452,-0.687) | 2.972 | -0.483 | net effect |
| M012 | (ROE) Return on equity | (1.707,2.984,5.641) | (-0.334,-0.441,-0.67) | 3.329 | -0.472 | net effect |
| M013 | (EPS) Earnings per share | (1.058,2.122,4.332) | (-0.178,-0.234,-0.356) | 2.409 | -0.251 | net effect |
| M021 | Revenue growth rate | (1.19,2.297,4.597) | (-0.196,-0.26,-0.394) | 2.595 | -0.277 | net effect |
| M022 | Net profit growth rate | (1.233,2.352,4.682) | (-0.313,-0.414,-0.63) | 2.655 | -0.443 | net effect |
| M023 | Sustainable growth ratio | (1.031,2.085,4.276) | (-0.141,-0.185,-0.282) | 2.369 | -0.198 | net effect |
| M031 | Economical risk | (1.259,2.388,4.735) | (-0.241,-0.318,-0.484) | 2.693 | -0.340 | net effect |
| M032 | Financial risk | (1.27,2.401,4.755) | (-0.338,-0.448,-0.68) | 2.707 | -0.479 | net effect |
| M033 | Systematic risk(beta) | (1.203,2.314,4.623) | (-0.117,-0.153,-0.234) | 2.613 | -0.164 | net effect |
| M041 | (MV/BV) Market value to book value | (1.173,2.272,4.559) | (-0.155,-0.204,-0.31) | 2.569 | -0.218 | net effect |
| M042 | (P/E) Price to income ratio | (1.138,2.228,4.492) | (-0.141,-0.186,-0.283) | 2.521 | -0.199 | net effect |
| M043 | (DPS/EPS) Divided ratio | (1.123,2.176,4.414) | (-0.193,-0.285,-0.433) | 2.472 | -0.299 | net effect |
| M051 | Exchange rate | (1.228,2.318,4.63) | (0.854,1.107,1.682) | 2.623 | 1.187 | net cause |
| M052 | Bank interest rank | (1.15,2.214,4.472) | (0.932,1.211,1.839) | 2.513 | 1.298 | net cause |
| M053 | Inflation rank | (1.101,2.119,4.326) | (0.905,1.265,1.921) | 2.416 | 1.339 | net cause |

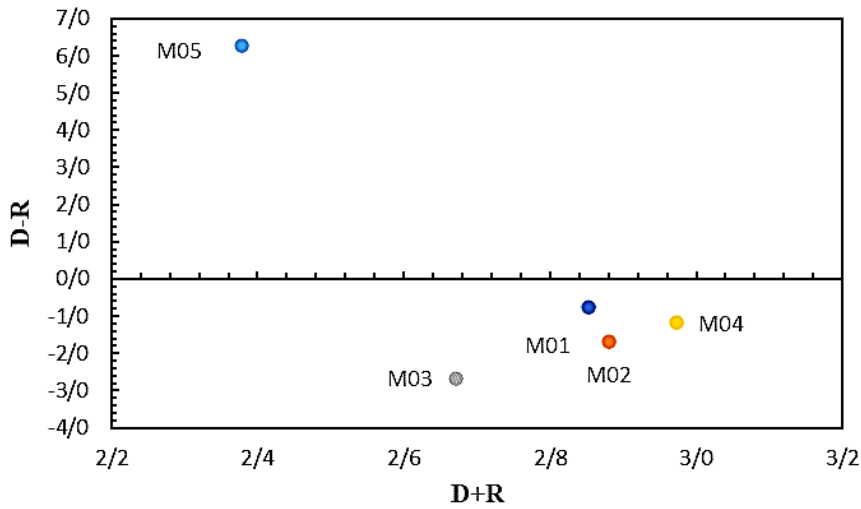

**Fig 1**. The causal diagram of the crisp data

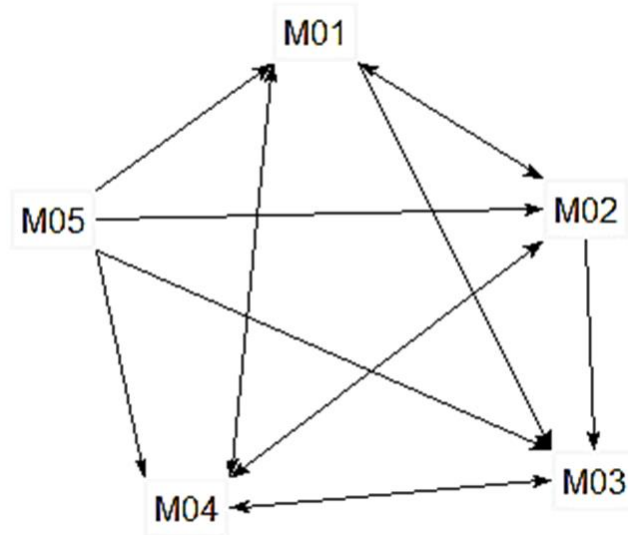

**Fig 2.** The Impact Relationship Map (IRM)

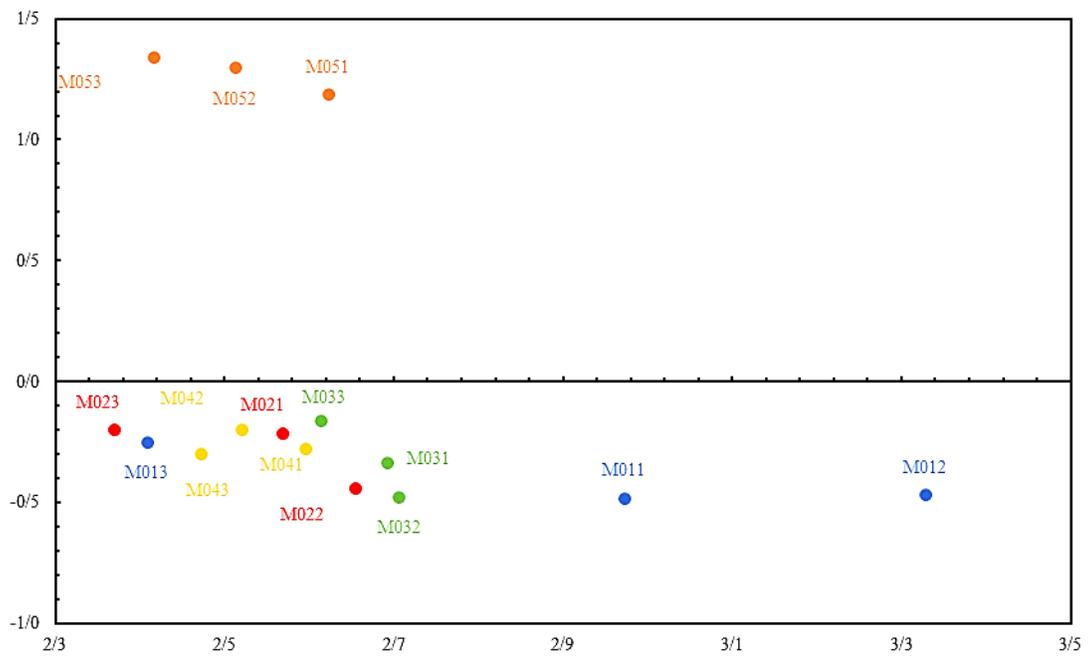

**Fig 3**. The causal diagram of the crisp data

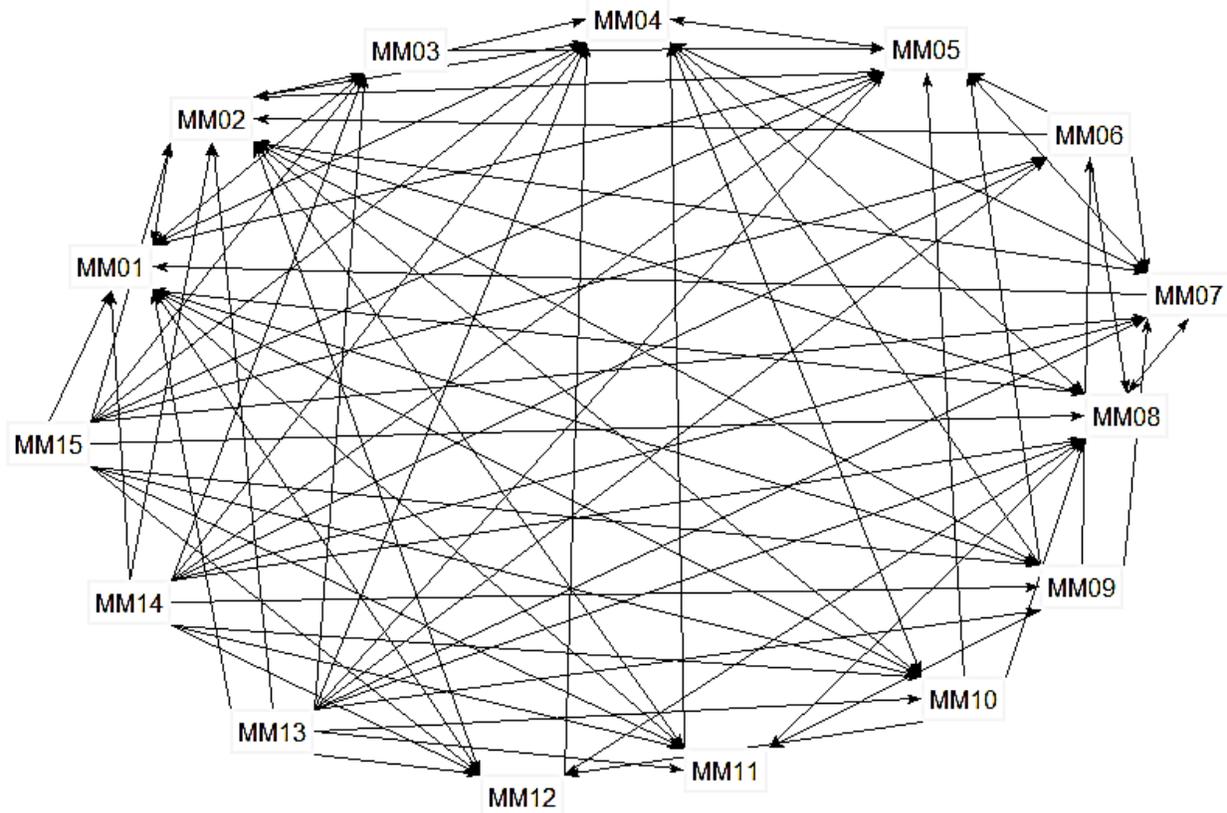

**Fig 4.** The impact relationship map of sub-impediments.

*1-3 Summary of results:*

It is shown in Tables 8 and 9 by each ranking.

Table 8. Results of sub-criteria ranking with DEMATEL

| Rankings | Criteria |
|---|---|
| 1 | Sub-criterion of Net Profit Growth Rate (GNP) |
| 2 | Sub-criterion of market-value-to-book value ratio (MV / BV) |
| 3 | Sub-criterion of standardization of methods (SP) |
| 4 | Sub-criterion of Asset in Return (ROA) |
| 5 | Sub-Criterion of Market Risk (BETA) |
| 6 | Sub-criterion of Systematic risk (RF) |
| 7 | Sub-criterion of Net Profit Margin (NPM) |
| 8 | Sub-criterion of Financial Risk Sub-Criterion (RM) |
| 9 | Sub-criterion of the Growth revenue ratio (GE) |
| 10 | Sub-criterion of net profit Growth Rate (GEPS) |
| 11 | Sub-Criterion of Management Commitment (MM) |

Table 9. Results of criteria ranking results with DEAMATEL

| Rankings | Criteria |
|---|---|
| 1 | Profit-making criterion (Profit) |
| 2 | The criterion of Methods and Operations Management (MPO) |
| 3 | Market benchmark |
| 4 | Risk criterion |
| 5 | Growth criteria |

Due to the model's appropriateness and reliable credibility in selecting the optimal stock portfolio, it is suggested that portfolio managers in companies, investment funds, and capital owners use this to select the optimal stock in the Toronto Stock Exchange. As illustrated in the Fig 3, the earnings per share growth rate are highly prioritized. Thus, it is suggested that investors consider this criterion while choosing a stock portfolio. The net profit margin rate has been of great importance in-stock selection by experts, so it can be recommended that investors utilize this criterion as an adequate criterion when choosing a portfolio and apply it to their decisions. Since systematic risk is of high relative importance for stock selection, it is highly recommended to portfolio managers and investors in the stock market to examine the existing risks, especially systematic risk, which is of great significance in all models of stock selection and increases the return on their portfolio by considering this variable.

Standardization of methods has been one of the essential issues for stock selection from the viewpoint of experts. Thus, standard methods or models developed for stock selection are generally not used or utilized trivially in the country. Hence, investors and policymakers in the Toronto Stock Exchange need to pay more attention to existing models to select the stock portfolio. Management commitment is one of the criteria affecting stock returns. According to experts, it has been of great importance in portfolio selection, so investors must consider and select this variable to observe each company's stock on the stock exchange at Toronto Stock Exchange.

## 4. Discussion

The active and robust capital market is consistently recognized as one of the features of developing countries internationally. In developed countries, plenty of investments are made via financial markets. The dynamic participation of people in the stock market guarantees the nourishment of the capital market and the country's sustainable development. However, investors' main problem in these markets is finding suitable investment securities and forming the optimal stock portfolio. The investment process in a coherent state requires the analysis of the primary nature of decisions in investment. In this case, the activities related to the decision-making process are analyzed, and the essential factors in the environment of the investors that affect their decisions are examined as well. Every day, extensive efforts are made to improve the stock analysis methods in the world's financial markets. The effort to improve stock analysis methods, especially in markets where the number of stocks is very high, has contributed to the emergence of new methods that, along with the previous methods, seek to find an answer to the desire to maximize one's profit in the financial markets. However, these mentioned methods could not adapt to the conditions of the capital market in Iran and have a significant effect on the choice of investors. On the other hand, the clarifications made in the last few years in the stock exchange have led to access to a large amount of technical information.

The proper use of this information is not possible for ordinary people and needs to utilize the opinions of financial experts. The information and other influential factors have made individual decisions to choose the right stock portfolio a problematic issue to the extent that most people base their criteria for deciding on stock selection on the volume of buying and selling queues, news, and rumors heard in the market and issues like this have reduced. One of the controversial issues here is managing this massive amount of information and practical usage to improve the decision-making process. One key issue for investors in such markets is choosing shares on the stock exchange. Also, finding ways to help investors choose stocks in these markets is essential. One problem for investors is choosing a suitable benchmark for comparing different stocks. The optimal functioning of the capital market occurs when investors are active in the market with comprehensive information and full awareness so that this market can create value for individuals and help the production cycle at the macroeconomic level. If the investor makes a rational decision when choosing stocks, he can achieve the desired return.

An important factor that can help investors in the optimal selection of stocks is to observe the criteria approved by financial experts and investors. The main point in stock investing is that decision-making is neither linear nor one-dimensional. However, a successful decision-maker examines the decision subject from different viewpoints and utilizes several criteria jointly and simultaneously. In addition to investigating the factors affecting it, selecting the best priority option has been mentioned.

## 5. Conclusion

Multi-criteria decision-making approaches make this possible for the investor as the algorithm is highly efficient and based on mathematical logic, compatible with human thinking and mental processes. Thus, the current study investigates the key factors influencing the stock exchange selection on the DEMATEL method Multi-Criteria Fuzzy decision-making approaches regarding the development of the Gordon model. As an essential tool of the capital market, the stock market plays a unique part in the economy. Stock exchange connotes an organized and formal capital market in which the purchase and sale of shares of companies or government bonds, or authentic private institutions are made under special rules and regulations. An essential feature of the stock exchange in the private sector is savings and liquidity collection centers to finance long-term investment projects. On the other hand, it is an official and reliable reference that shareholders of stagnant savings can find a relatively suitable and safe place to invest and use their surplus funds to invest in companies or attain a confident and guaranteed profit by buying bonds from governments' authentic companies. Based on the results obtained from the ranking of sub-criteria, criteria affecting the selection of stocks in the Toronto Stock Exchange using experts and Excel software are described below.

Regarding the model's appropriate and reliable validity in choosing the optimal stock portfolio, it is recommended that portfolio managers in companies, investment funds, and capital owners use the model to select stocks in the Toronto Stock Exchange optimally. It is also recommended for future works that researchers add new criteria such as liquidity, floating shares, stop symbol duration, risk-free interest rate, and other financial ratios to the model. It can increase the model's efficiency. In addition, using other decision-making techniques, such as Electra-Vicor, can provide more favorable results, and various scenarios can be examined.

**Funding:** This research received no external funding.

**Data Availability Statement:** There is no available data.

**Conflicts of Interest:** The authors declare no conflict of interest.